Effect of Bi-substitution on Structural Stability and Improved Thermoelectric Performance of p-type Half-Heusler TaSbRu: A First-principles Study

Enamul Haque*[1], Mostafizur Rahman[2], and Parvin Sultana[2]

[1]EH Solid State Physics Laboratory, Longaer, Gaffargaon, Mymensingh-2233, Bangladesh

[2]Department of Physics, Bangladesh University of Engineering and Technology(BUET), Dhaka-1000, Bangladesh

*Email: enamul.phy15@yahoo.com, enamull@mailaps.org

**Abstract**

Recently, Fang et al. have predicted a high ZT of 1.54 in TaSbRu alloys at 1200 K from first-principles without considering spin-orbit interaction, accurate electronic structure, details of phonon scattering, and energy-dependent holes relaxation time. Here, we report the details of structural stability and thermoelectric performance of Bi-Substituted p-type TaSbRu from first-principles calculations considering theses important parameters. This indirect bandgap semiconductor ($E_g$=0.8 eV by TB-mBJ+SOC) has highly dispersive and degenerate valence bands, which lead to a maximum power factor, 3.8 mWm$^{-1}$K$^{-2}$ at 300K. As Sb-5p has a small contribution to the bandgap formation, the substitution of Bi on the Sb site does not cause significant change to the electronic structure. Although the Seebeck coefficient increases by Bi due to slight changes in the bandgap, electrical conductivity, and hence, the power factor reduces to ~3 mWm$^{-1}$K$^{-2}$ at 300K (50% Bi). On the other side, lattice thermal conductivity drops effectively to 5 from 20 W/m K as Bi introduces a significant contribution in the acoustic phonon region and intensify phonon scattering. Thus, ZT value is improved through Bi-substitution, reaching 1.1 (50% Bi) at 1200 K from 0.45 (pure TaSbRu) only. Therefore, the present study suggests how to improve the TE performance of Sb-based half-Heusler compounds and TaSbRu (with 50% Bi) is a promising material for high-temperature applications.

1. **Introduction**

Thermoelectric (TE) devices operating at high temperatures need to be constructed by using the materials with a high melting temperature and high TE conversion efficiency. Only a few materials have been discovered in the past decades [1–5], that exhibit not only high-temperature stability but also relatively high TE performance. Half Hesuler compounds, one of such type phase, have thus been extensively explored in recent years [5–10]. These compounds have an excellent electronic structure (highly dispersive and large degeneracy, that can lead to high power factor) and relatively high TE performance [11]. The Half-Heusler compounds, with 18 valence electrons (closed-shell configuration), are semiconductors and the bandgap can be either narrow (usually) or wide. The TT'S' (T=Ti, Zr, Hf; T'=Co, Ni, Zn; and S'= Sn, Sb) compounds and their alloys widely studied and found to be promising materials for high-temperature TE applications [12–14]. The dimensionless quantity, ZT (=$S^2\sigma T/\kappa_{tot}$; where S, $\sigma$, T and $\kappa_{tot}$ are the Seebeck coefficient electrical conductivity, absolute temperature, and total thermal conductivity (lattice plus electronic contribution, respectively)[15–17]), called figure of merit, describes the performance of material as thermoelectric. Half-Heusler phases usually possess a high power factor ($S^2\sigma$), but on the other side, high thermal conductivity. Therefore, the practical applications of these materials are limited and needed to optimize their performance. There have been developed many techniques for this, such as nanostructuring and alloying [18,19]. For example, the substitution Ti on Nb site in FeNbSb has been found experimentally to enhance TE performance (ZT=0.8-1.5 at 1100-1200K) substantially [20–23]. The alloying causes the intensification of phonon scattering and hence reduces the lattice thermal conductivity effectively. However, we need to be careful about the choice of dopant such that the dopant does not change the electronic structure significantly and, hence the power factor.

TaRuSb, a Half-Heusler compounds synthesized about 20 years ago [24], has been recently reported from theoretical studies that it is a promising TE material [25–27]. Yang et al. reported the high power factor of TaSbRu (2.8 mW/m k2) by using PBE functional without SOC and constant relaxation time 10-14 s) at 300 K [25], but Fang et al reported it to be 9 mW/m k2 [26]. The latter group of authors also reported ZT=1.54 in 8% Ge substituted TaSbRu at 1200K, stating the low band effective mass caused such a high ZT [26]. Moreover, they reported that Ge is the best dopant among Ge, Sn, Pb [26]. Generally, heavier elements have been found to reduce lattice thermal conductivity effectively [28,29]. They used a simple Slack equation (which considers acoustic phonons scattering only) for lattice thermal conductivity calculations, PBE functional without spin-orbit coupling (SOC), and energy independent relaxation time for carrier transport calculations. However, Kaur and Kumar have reported the much higher PF value, 21 mW/m k2, and lower lattice thermal conductivity (4.19 W/m K at 300 K (while it reported being 17.9 in the Ref. [26])) in the same compound. They also reported wrong phonon dispersion. Furthermore, the authors of the latter two groups reported different carrier relaxation time of TaSbRu (1 fs in Ref.[27] and 37 fs in Ref.[26] ) by using the same method. Overall, there are large discrepancies between the reported theoretical TE properties of TaSbRu in the literature.

Here, we report the electronic structure of TaSbRu by mBJ+SOC, carrier transport coefficients by using energy-dependent relaxation time and lattice thermal conductivity from IFCs by solving phonon Boltzmann transport equation, and other related parameters. We also present the effect of Bi- substitution in TaSbRu and report substantial improvement of thermoelectric figure of merit, as Bi reduces the lattice thermal conductivity dramatically. Moreover, we explain the possible reason behind the large discrepancies of TE performance reported in the literature. Our study suggests that TaSbRu alloys are promising materials for thermoelectric devices operated at high temperatures.

2. Computational Details

The electronic structure calculations were performed by using a full potential linearized augmented plane wave (LAPW) method in wien2k [30]. First, we performed full structural relaxation by minimizing energy (and force for variable positions, where it is relevant). In all calculations, we set strict convergence criteria, energy, and charge convergence to $1 \times 10^{-6}$ eV and 0.001e, respectively. The exchange-correlation term was treated by using the generalized gradient approximation (GGA) [31] with PBEsol [32] setting and BZ integration was performed with $10 \times 10 \times 10$ k-point. Although the PBEsol functional underestimate the lattice parameters slightly, it has a good track of record for the calculation of correct phonon properties as compared to experimental results. We modeled Ta, Sb Bi, and Ru muffin sphere with radii (Rmt) 2.35, 2.15, 2.25 Bohr, and used valence band and conduction band separation energy 7 Ry, and kinetic energy cutoff RmtKmax 7. As GGA underestimates the bandgap (about 50% by experimental value),

Tran-Blaha modified Becke Jonshon potential (TB-mBJ [33–35], commonly known as mBJ) was used for electronic structure computation utilizing a finer k-mesh 34 × 34 × 34 (34 × 34 × 24 for tetragonal). The calculated energy eigenvalues and other parameters were then fed into a modified BoltzTraP code to calculate the transport coefficient. In all these calculations, we considered spin-orbit interactions. As original BoltzTraP calculates transport coefficients within constant relaxation time approximation (cRTA) [36], we need to calculate electron-phonon dynamical matrix to calculate relaxation time in the modified code. For this, we used projector augmented plane wave method (PAW) and density functional perturbation theory (DFPT), in Quantum Espresso (QE) [37], using 50 Ry cutoff energy for wavefunctions, 888 (884 for tetragonal symmetry) k- and q-point, with the same PBEsol functional, and energy threshold $10^{-10}$ Ry setting and projector-augmented wave (PAW) pseudopotentials (provided in PSlibrary 1.0 [38]). The obtained dynamical matrix was also used for the phonon dispersion and density of states calculations. The required average electron-phonon dynamical matrix was then obtained within the energy bins based method in EPA code [39]. In this calculation, we used CBM and VBM energy, energy interval 0.7 eV, and 6 energy bins. The output was used to calculate hole relaxation time in the modified BolzTraP. Note that the structural relaxation was performed again in QE and almost the same lattice parameters were obtained within computational accuracy.

The lattice thermal conductivity and its related parameters were computed by using the finite displacements approach as implemented in Phono3py [40,41]. To calculate second and third-order force constants (IFCs), we created 222 supercell containing 24, 96, 48 atoms (for pure, 25% Bi, and 50% Bi systems, respectively). The force calculations were performed in the QE by using the same setting except for k-point (2 × 2 × 2 k-point in that case). The obtained IFCs were then used to compute lattice thermal conductivity by solving the phonon Boltzmann transport equation within the cRTA, by using 16 × 16 × 16 q-point. This method has been extensively used in the recent decade to compute lattice thermal conductivity accurately [41–43].

3. Results and Discussion

Half-Heusler TaSbRu crystallizes in a face-centered cubic structure (F-43m, #216 [24]) with four formula unit ( 3 and 12 atoms in the primitive cell and conventional cell) as shown in the Fig. 1. (a). In the structure, Ta, Sb, and Ru atoms occupy 2a( 0, 0, 0), 2c(¼, ¼, ¼), and 2b (½, ½, ½)

Wyckoff positions respectively. The bond length between Ta-Sb and Ru-Sb bond length is 3.07Å. The atomic arrangement suggests that phonons can propagate along with Sb-Ta-Sb and Sb-Ru-Ta-Sb directions easily. If we replace Sb (0.25, 0.25, 0.75) from the conventional unit cell by Bi and apply symmetry, the symmetry is lowered to a simple cubic structure (P-43m, #215), as shown in Fig. 1. (b). In this structure, Ta, Sb, Bi and Ru atoms located at 4e(….), 3d(1/2, 0, 0), 1a(0, 0, 0) and 4e. The phonons might scatter by Bi when it propagates q along with Ta-Sb-Sb directions.

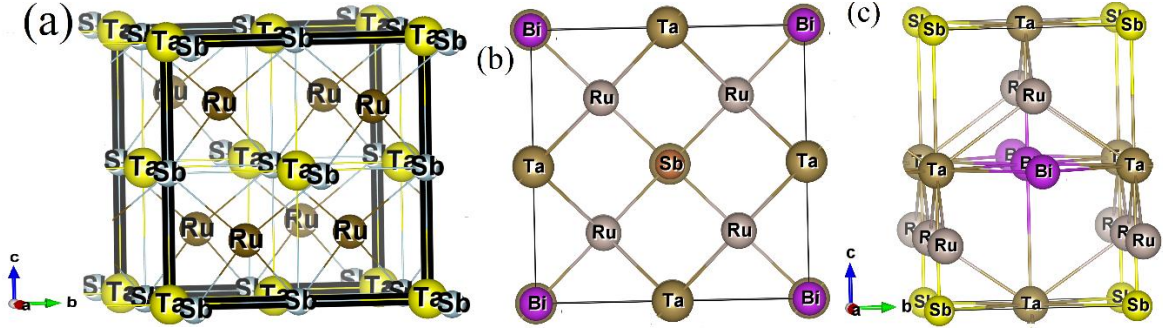

Fig. 1. Crystal structure of: (a) face-centered cubic TaSbRu, (b) simple cubic 25% Bi substituted, and (c) simple tetragonal 50% Bi substituted TaSbRu.

The bond length among the constituent atoms is slightly reduced (2.97 Å) as compared to the original structure. The atomic arrangement is further modified when we replace another Sb (0.25, 0.75, 0.25) by Bi (50% Bi) as illustrated in Fig. 1. (c). The symmetry is also much lowered to tetragonal structure (P-4m2, #115) and reduced the number of atoms to 6 from 12. In this symmetry, Ta, Sb, Bi and Ru atoms situated at 2e(1/2, ½, 0), 1a(0, 0 , 0), 1b(1/2, ½, 0), and 2f (1/2, ½, ..), respectively. In this atomic arrangement, the phonons may have strong scattering along with Sb-Ru-Ta directions by Bi. Therefore, the lattice thermal conductivity might be reduced significantly in this symmetry. Our calculated lattice parameters, formation energy, elastic constants and moduli, Debye temperature, sound velocity, Pugh's ratio, Poisson ratio, melting temperature, and other available experimental and theoretical values are listed in Table I.

Table I: Computed fully relaxed lattice parameters (a, c), Formation energy (H), Elastic constants ($c_{ij}$) and moduli (B, G), Debye temperature ($\theta_D$), sound velocity (v), Pugh's ratio (B/G), Poisson ratio (v), melting temperature (Tm), and other available experimental and theoretical values (enclosed in the bracket).

| Parameters | TaSbRu | 25% Bi | 50% Bi |
|---|---|---|---|
| a (Å) | 6.11,(6.19 PBE [26], 6.024[27], 6.097 LDA [44], 6.13 EXP [24]) | 6.14 | 4.36 |
| c (Å) | - | - | 6.1691 |
| H (eV/atom) | -0.4 | -0.8 | -0.53 |
| $c_{11}$ (GPa) | 301 (306 LDA [44]) | 293 | 277 |
| $c_{12}$ (GPa) | 150 (157 LDA [44]) | 146 | 152 |
| $c_{44}$ (GPa) | 74 (122 LDA [44]) | 67 | 61 |
| $c_{33}$ (GPa) | - | - | 286 |
| $c_{13}$ (GPa) | - | - | 143 |
| $c_{66}$ (GPa) | - | - | 72 |
| B (GPa) | 201 (179 PBE [26], (203 by EOS [44]) 206 LDA [44]) | 195 | 191 |
| G (GPa) | 74 (71 PBE [26], 103 LDA [44]) | 70 | 65 |
| Y (GPa) | 199 (259 LDA [44]) | 187 | 176 |
| B/G | 2.69 (2.05 LDA [44]) | 2.8 | 2.9 |
| $v$ | 0.33 (0.29 LDA [44]) | 0.34 | 0.35 |
| $v_l$ (m/s) | 5060 (4854 PBE [44]) | 4859 | 4684 |
| $v_t$ (m/s) | 2523 (2472 PBE [44]) | 2393 | 2270 |
| $v_a$ (m/s) | 2831 (2770 PBE [44]) | 2687 | 2551 |
| $\theta_D$ (K) | 316 (305 PBE [26], 392 LDA [44]) | 298 | 282 |
| $T_m$ (±500 K) | 2476 | 2422 | 2382 |

The lattice parameters of pure TaSbRu fairly agrees with the experimental value (only underestimate about 0.3%, which is within the limitation of GGA), while PBE functional overestimates (underestimates) the lattice parameters by 1% (0.62%, 1.8%). Our calculated values of other parameters by using PBEsol functional are consistent with the reported theoretical values (by using PBE functional). The negative value of formation energy indicates that studied compound and alloys are energetically favorable to form in the laboratory. However, these materials need to dynamically stable for this. We will discuss this issue in section 3.3. The positive elastic constants and bulk moduli satisfy the stability conditions as described in Ref. [45]. Therefore, the systems under study are mechanically stable. The bulk modulus is slightly reduced by Bi-substitution. Thus, the mechanical resistance of these materials is also reduced. The Debye temperature is significantly reduced by alloying through Bi, which indicates the intensification of phonon scattering. The systems are still ductile and ductility is slightly improved. The melting temperature is a very important parameter for high-temperature TE materials. Our estimated melting temperature of the compound and alloys are high enough to operate them above 1200 K. The parameters listed in table I show an overall good agreement with PBE results but largely differ from that obtained by LDA along a certain direction of propagation ($c_{44}$).

### 3.1. Electronic Structure

Half-Heusler compounds exhibit some interesting features in their electronic structure, such as Half-metallicity, high band degeneracy, and dispersive bands, etc. Our spin-polarization calculations (not presented here) suggest that the materials under consideration are non-magnetic. The electronic shell configuration of TaSbRu (Ta(5d3 6s2), Sb(5s2 5p3), Ru(4d7, 5s1)) is closed, thus, it must be a semiconductor. The Bi (6s2 6p3) has the same electron configuration. Therefore, it should not cause an unpaired shell configuration. That is, the alloys should be still semiconductors upon Bi-substitution. Our computed electronic band structure and projected density of states of the compounds and its alloys are shown in Fig. 2. The CBM lies at the high symmetry point X ($\Gamma$ for 25% Bi and M for 50% Bi) and VBM at L ( R for 25% and 50% Bi substitution), suggesting an indirect nature of bandgap. For TaSbRu, the VBM is twofold degenerate bands and each band has four extreme valleys (number of degeneracy (Nv)=8) and CBM is onefold degenerate bands with three extreme valleys ($N_v$=3). Furthermore, both CB and VB are highly dispersive. These features of the electronic structure may be conducive for holes transport. The computed values of bandgap by using mBJ+SOC of the studied compound and alloys are listed in Table II.

Table II: Calculated bandgap ($E_g$ (eV)) of TaSbRu and its alloys with other available theoretical values.

| Materials | mBJ+SOC | PBE |
|---|---|---|
| TaSbRu | 0.80 (0.88 by mBJ [44]) | 0.653 [26], 0.655 [25], 0.55 [27], (0.65 LDA [44]) |
| 25% Bi | 0.81 | - |
| 50% Bi | 0.83 | - |

Note that our calculated PBEsol bandgap (0.64 eV) of TaSbRu fairly agrees with the other reported theoretical values (by using PBE functional). Generally, PBE functional underestimates the experimental value by ~50%. However, TB-mBJ potential correction to the electronic structure can produce bandgap close to the experimental values. As Bi is isoelectronic to Sb, the substitution of Bi does not cause a significant change in the bandgap value, the gap is slightly increased from 0.8 to 0.83 eV (less than 5%) by 50% Bi substitution within the computational accuracy. From the projected density of states (Fig. 2. (b, d, f), Ta-5d, and Ru-4d have a dominated contribution to the CBM and VBM and formulate the bandgap. Sb-5p/Bi-6p has a minor contribution to the VBM.

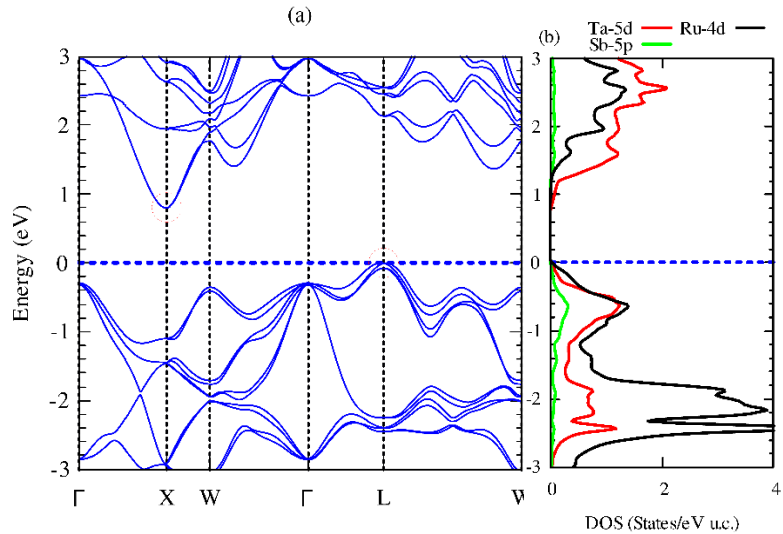

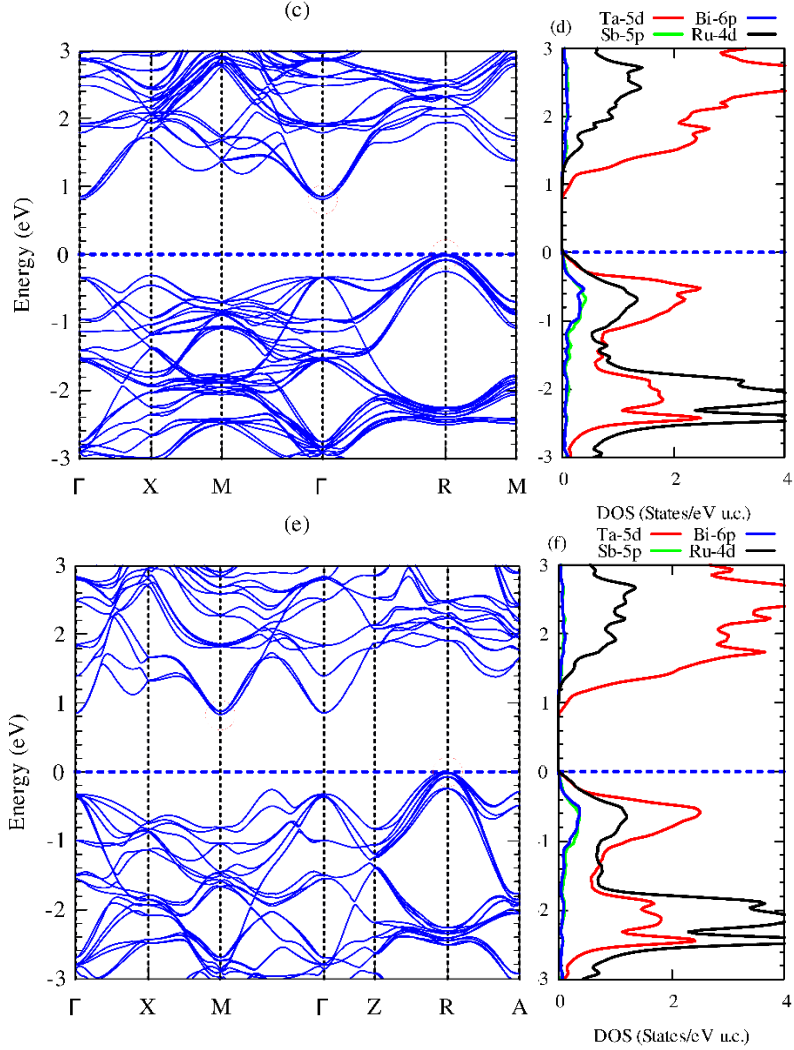

Fig. 2. Electronic band structure and partial density of states of: (a-b) pure, (c-d) 25% Bi substituted, and (e-f) 50% Bi substituted TaSbRu, respectively. The blue dashed lines at zero energy represent the Fermi energy. The open circles indicate the minima of the conduction band (CBM) and maxima of the valence band (VBM).

However, the dispersive character of CBM is reduced. The VBM for 25% and 50% Bi alloys are twofold degenerate bands and each band has three extreme valleys ($N_v$=6). It is not clear why isoelectronic Bi causes such changes to the electronic structures. This remains an open question. Further studies are required to address this question. The changes in band degeneracy and dispersive nature of VBM may cause the reduction of electrical conductivity and hence, the power factor. Note that the change of dispersive character of 25% is large as compared to TaSbRu but small as compared to 50% Bi. The band edges much sharpen in the case of 50% Bi substitution, indicating a high Seebeck coefficient and low electrical conductivity in this case as compared to others.

### 3.2. Lattice dynamics

The dynamical stability is a very important criterion for a compound/alloys to synthesize in the laboratory. The attempt for synthesizing a compound/alloys may fail due to the dynamical instability. For this reason, we have computed the phonon dispersion and density of states to verify the stability of TaSbRu and its alloys. The results are presented graphically in Fig. 3. Although the phonon dispersion of TaSbRu fairly agrees with the reported results in Ref. [26] by PBE and Ref. [46] by PBesol, but it largely differs from the reported one in Ref. [27]. In Ref. [27], the highest optical mode frequency is ~4 THz (~8.8 THz in our case) and a small gap exists between acoustic and optical phonons. Although the authors of the article used the same method (QE) as we, it is unexpected to get such large discrepancies. When we use the wrong atomic mass ( it must be provided in a.m.u unit) in the QE input, such type wrong phonon dispersion will be resulted. Also we must have to use converged k/q-points and cutoff energyfor accurate phonon calculations. However, a small gap between acoustic and optical phonons from X to L of TaSbRu exists ~around 4 THz energy, which will be conducive for phonons. This suggests a high lattice thermal conductvity in TaSbRu is probable. In this region, Ta and Sb has largest contributions to the phonons while higher optical phonons arise mostly from Ru and Sb. Since the higher optical phonons generally has negligible contributions to heat conduction, we need to concentrate on the acoustic phonons to reduce lattice thermal conductivity. As we mentioned in the previous section that Ta and Ru formulates the bandgap, these sites are not suitable to replace by other elements. The replacement one Sb (25%) by Bi causes loering the acoustic phonon energy and remove the small gap. Along these high symmetry points except Γ. Now the phonon density of states is dominated by combined contribution of Sb and Bi. The gap at Γ is further narrowed by the replacement of another Sb with Bi and phonon energy significantly lowered. In this case, the Bi has domindated contribution to the acoustic phonons. This suggest us the lattice thermal conductvity should be dramatically reduced.

In all of our calculations, we consider LO-TO splitting effect by calculating macroscopic dielectric constant and effective charges. The dielectric constant is increased by a few percent (from 19.5 to 21.4) due to the Bi substitution. Interestingly, such change causes the presence of LO-TO splitting in the case of 50% Bi at $\Gamma$-point as shown in Fig 3. (e). We note that all the phonon modes have positive energy and thus, the compound and alloys under consideration are thermodynamically stable, suggesting the possibility of synthesizing them in the laboratory. For TaSbRu, it should be because the compound has been synthesized already.

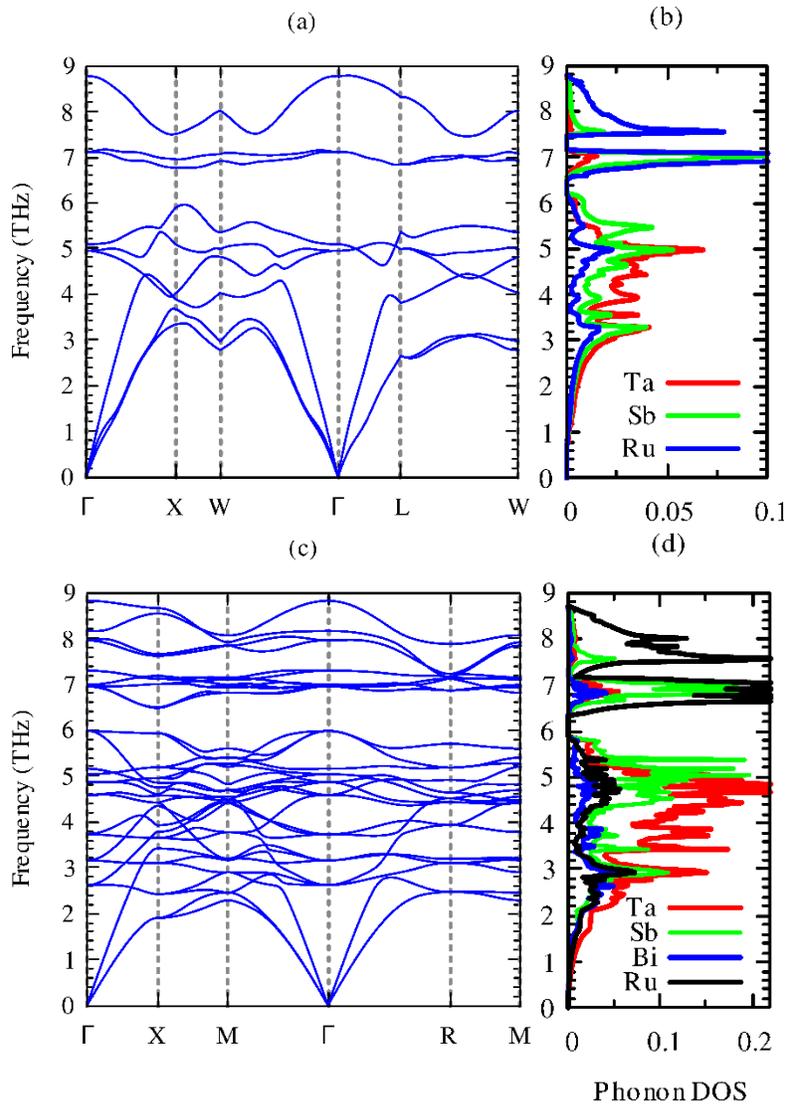

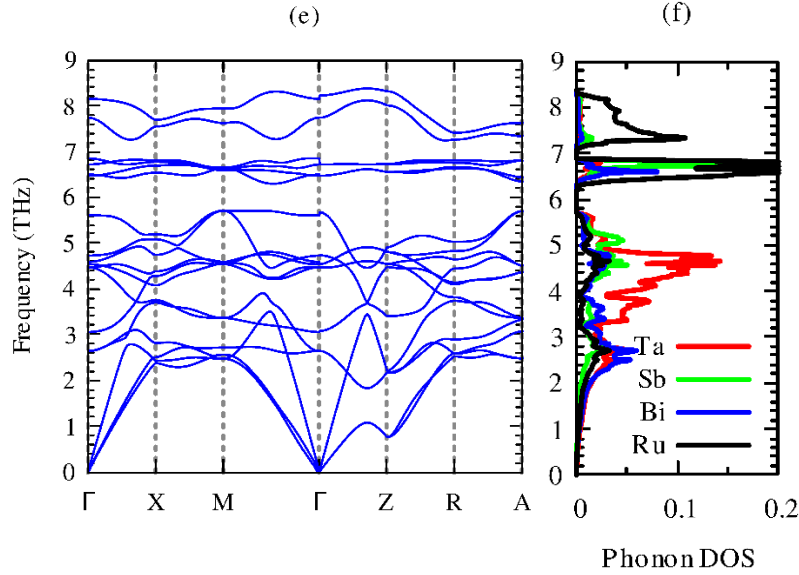

Fig. 3. Phonon dispersion relations and partial phonon density of states of: (a-b) pure, (c-d) 25% Bi substituted, and (e-f) 50% Bi substituted TaSbRu, respectively.

Now let us discuss the phonon lifetime of the studied materials at 1200 K calculated from second and third-order IFCs as shown in Fig. 4. The maximum phonon lifetime of TaSbRu is 8 ps, while it almost two times smaller in 25% Bi substituted alloy. It is further reduced (~6 times smaller than that of TaSbRu) by 50% Bi substitution as shown in Fig. 4. (c). Therefore, the Bi-substitution not only lowers the acoustic phonon energy but also it shorters the phonon lifetime.

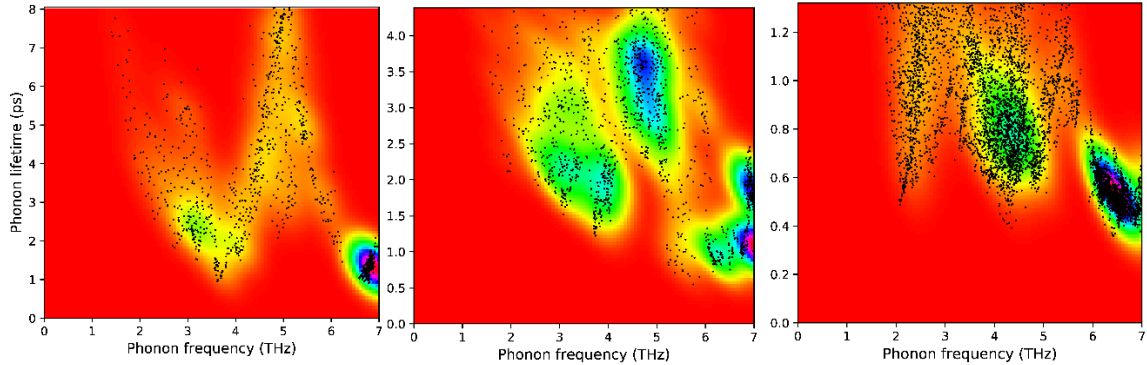

Fig. 4. Phonon lifetime at 1200K temperature of: (left panel) pure, (middle panel) 25% Bi substituted, and (right panel) 50% Bi substituted TaSbRu, respectively.

The computed phonon group velocity of TaSbRu and its alloys are shown in Fig. 5. Interestingly, the group velocity is slightly increased for some modes of phonons, especially for 25% Bi around 3.4 THz regions. But the acoustic phonons modes are significantly shifted to the lower energy. As higher optical phonons (see Fig. 6. (b) ), we do not present the data phonons with energy higher than 7 THz for clarity. In the case of 50% Bi alloy, the group velocity of optical phonons (especially above 4 Thz) is significantly decreased.

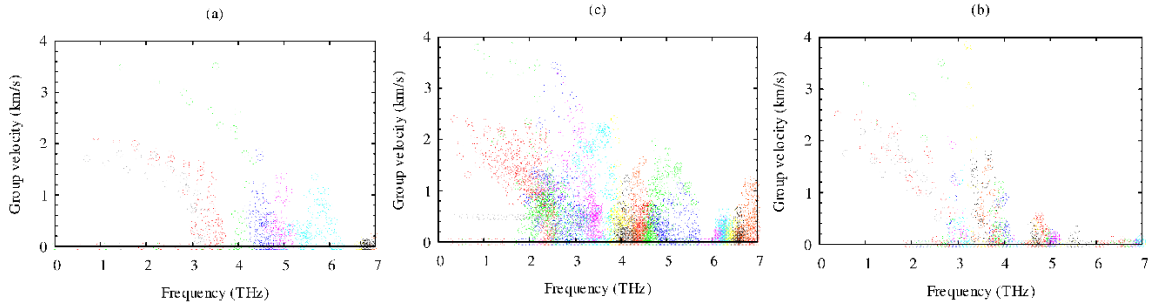

Fig. 5. Frequency dependent phonon group velocity of: (a) pure, (b) 25% Bi substituted, and (c) 50% Bi substituted TaSbRu, respectively.

Our computed values of mode Gruneisen parameter, an important quantity for the description of scattering, fo three systems are presented in fig. 6. A larger value of the Gruneisen parameter indicates high anharmonicity and hence, intense phonon scattering and vice versa. The Gruneisen parameter of TaSbRu is small as compared to that of low lattice heat-conducting materials. The mode Gruneisen parameter of 25% Bi alloy remains almost the same for some acoustic mode phonons but shifted to the lower energy. However, readers should note that the values of mode Gruneisen parameter for first acoustic phonon mode (grey colored open circles) of 50% Bi alloy are almost six times higher than that of the pure compound.

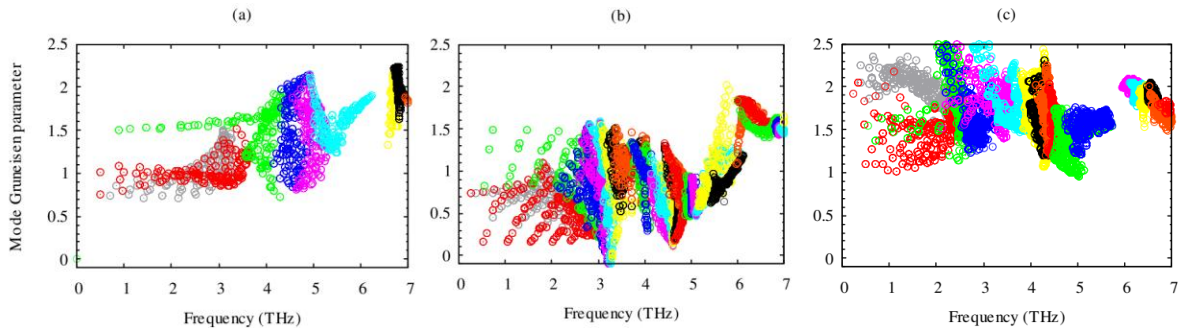

Fig. 6. Computed phonon mode Gruneisen parameter of: (a) pure, (b) 25% Bi substituted, and (c) 50% Bi substituted TaSbRu, respectively.

All these changes suggest that the phonon scattering has been intensified two-six times in Bi-substituted alloys as compared to the pure one. Now let us see the effect of these changes on the lattice thermal conductivity.

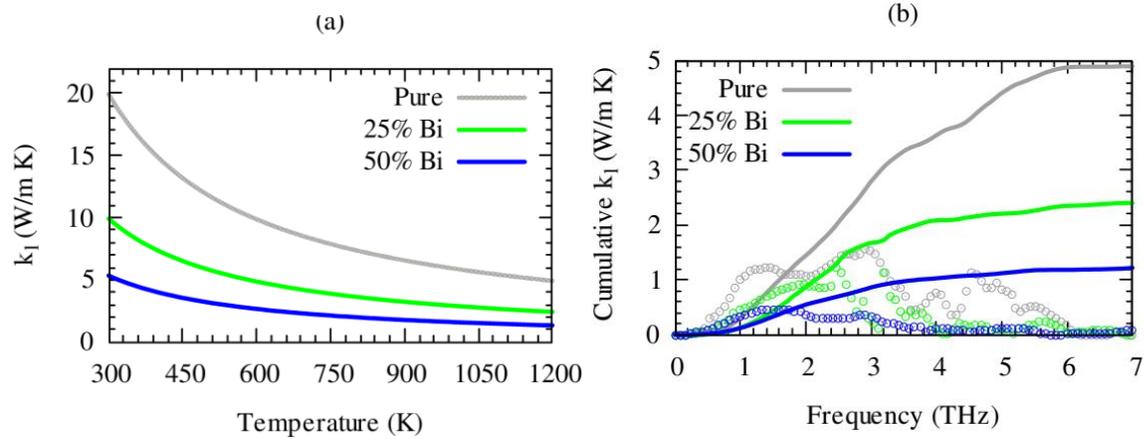

Fig. 6. (a) Lattice thermal conductivity at different temperatures and (b) cumulative lattice thermal conductivity and it's derivative at 1200K (open circles) of studied compound and alloys.

Fig. 6. shows the calculated lattice thermal conductivity at different temperatures (κl) (a), cumulative κl, and its derivative at 1200 K (b). From the figure, it is clear that the κl is two-four times smaller for 25-50% Bi alloys as compared to the pure one. This is expected results, as phonon scattering is largely intensified by Bi substitution. The substitution of heavier elements usually brings such type of change, as found in the case of many materials. But it is not clear why Ge is more effective to reduce κl of the same compound than Sn/Pb, as reported by using the Slack equation in the Ref. [26]. The Slack model considers only the acoustic phonon scattering and κl is directly proportional to the average atomic mass of the compound [47–49]. Furthermore, κl remains almost the same by Sn (17 W/m K), ~41% is reduced by Pb (10 W/m K) and 59% by Ge (7 W/m K), which is not in the expected order, as Ge, Sn, Pb are isoelectronic and only differ by atomic mass. Our computed room temperature lattice thermal conductivity and other available theoretical values are listed in Table III.

Table III: Room temperature lattice conductivity (W/m K) of TaSbRu and its alloys, with other available theoretical values

| Materials | PBEsol and pBTE | PBE and Slack |
|---|---|---|
| TaSbRu | 19.92 | 17.1, (20.9*, 15.58**, 4.19***) |
| 25% Bi | 9.91 | - |
| 50% Bi | 5.06 | - |

* κl from four exact anharmonic force constants; **by using quasi-harmonic Debye model; ***by PBE and pBTE in ShengBTE code.

Interestingly, κl of TaSbRu is close to that one reported one in the Ref. [26]. by considering acoustic phonon scattering only. However, about 10-20% of the lattice thermal conductivity comes from the optical phonons, which can be seen from the derivative of cumulative κl, as shown in Fig. 6. (b). That's why the Slack model underestimates κl by a few percent (also may be due to the use of overestimated lattice parameter). In the case of small dopant 25 % Bi (or 8% Ge in the Ref. [26].), optical phonons have still small (but not negligible) contribution to the lattice thermal conductivity. For these limitations of the equations used in Ref. [26] to calculate κl might be responsible for such an unexpected change of κl by Ge, or authors did serious mistakes in their calculations. Another important fact that κl reported in Ref. [27] is 4 times smaller than that of our values (also Ref. [26]). As the article reported wrong phonon dispersion, their reported κl might also have large uncertainty due to the wrong atomic mass used in their calculations. The HfCoSb, an isostructural compound with similar electronic structure, exhibits similar heat conduction (experimentally measured thermal conductivity at 400K is ~14 W/m K [50] ). However, the calculation of κl is very sensitive to the input parameters and large discrepancies have been found in both experimental and theoretical values for Half-Heusler compounds [51]. Therefore, it is very hard to say which one calculation/method accurately describes the κl.

### 3.3. Transport Properties

The calculation of hole relaxation time is performed with the help of modified BoltzTrap code [36,39], where it is computed by using the following equation [39]

$$\tau^{-1}(\epsilon, \mu, T) = \frac{2\pi\Omega}{g_s \hbar} \sum_v \{g_v^2(\epsilon, \epsilon + \bar{\omega}_v)[n(\bar{\omega}_v, T) + f(\epsilon + \bar{\omega}_v, \mu, T)]\, \rho(\epsilon + \bar{\omega}_v)$$
$$+ g_v^2(\epsilon, \epsilon - \bar{\omega}_v)[n(\bar{\omega}_v, T) + 1 - f(\epsilon - \bar{\omega}_v, \mu, T)]\rho(\epsilon - \bar{\omega}_v)\} \ldots \ldots (1)$$

where $\Omega$ stands for the unit cell (primitive) volume. Besides, $\hbar$ is the reduced Planck's constant, $v$ the index of phonon modes, $\bar{\omega}_v$ the averaged frequency of phonon modes, $g_v^2$ the averaged of the electron-phonon matrix, $n(\bar{\omega}_v, T)$ the Bose-Einstein distribution function, $f(\epsilon + \bar{\omega}_v, \mu, T)$ the Fermi-Dirac distribution function, $g_s = 2$ the degeneracy of spin, $\epsilon$ the energy of electrons, and $\rho$ the DOS per unit volume and energy. Readers are suggested to consult with the Ref. 45 for further details.

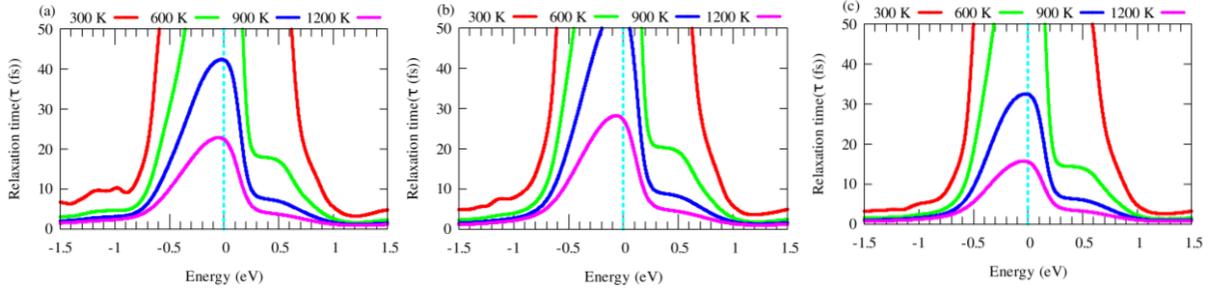

Fig. 7. Carrier relaxation time as a function of energy at different temperatures of (a) pure, (b) 25% Bi substituted, and (c) 50% Bi substituted TaSbRu, respectively. The dashed lines at zero energy represent the Fermi level of 300 K.

Fig.7. shows the energy-dependent carrier relaxation time at three different temperatures. The conduction bands (electrons) have two times longer relaxation time than that of valence bands in all cases considered here. Because the CBs are highly dispersive as compared to the VBs. This indicates high electrical conductivity and hence, high power factor in n-type TaSbRu and its alloys. However, all these systems intrinsically p-type, so we do not discuss the n-type system's properties. The room temperature holes $\tau$ of pure TaSbRu is 65 fs, which is about two times larger than that of reported (37.52 fs) in the Ref. [26] (two times smaller than that of reported in Ref. [27] (100 fs)) considering acoustic phonons and carrier scattering only. Although in these two references, authors used the same method, a large discrepancy exists between their reported values. The substitution of Bi causes the intensification of carrier scattering and hence, reduces the $\tau$ value.

The temperature has a similar effect on τ. In all cases, τ falls rapidly around the band-edges. Such type of effect can be explained by using the following equation [39]:

$$\tau^{-1} \sim g^2(\epsilon)\rho(\epsilon) \ldots \ldots \ldots (2).$$

The above equation suggests that τ is inversely proportional to the holes DOS (ρ) per unit energy and unit volume, while the electron-phonon dynamical matrix elements (g) exhibit a weak dependency on the energy of holes.

Our calculated Seebeck coefficient (S) and electrical conductivity (σ) of all systems are illustrated in fig. 7. At low temperatures, the S of pure TaSbRu is small and increased significantly with both temperature and Bi- substitution. Interestingly, the S in the high carrier concentration region of 50% Bi alloy remains almost the same as compared to that of the 25% Bi alloy. This is because 50% of Bi substitution has a negligible effect on dispersive character f VBM as compared to 25% Bi (see fig. 2 (c, e). On the other side, electrical conductivity shows the opposite trend as shown in Fig. 7. (b).

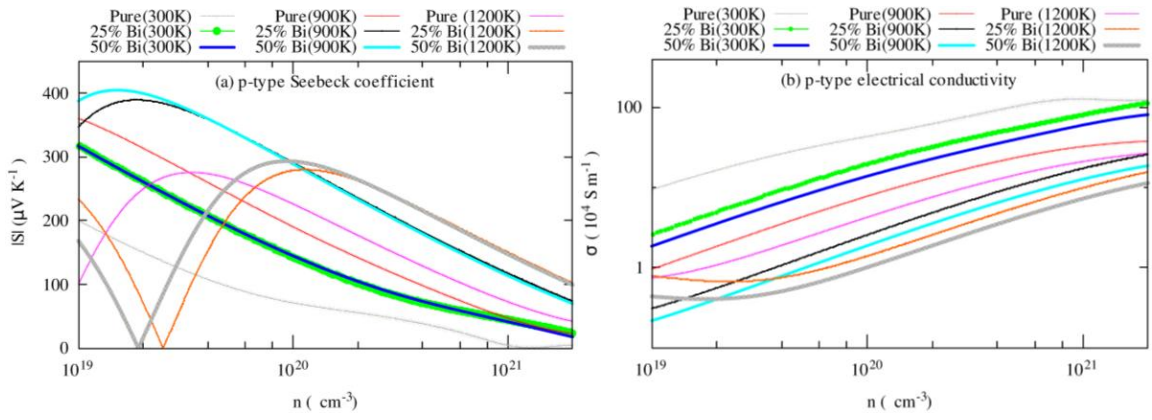

Fig. 7. (a) P-type Seebeck coefficient and (b) electrical conductivity of the studied compound and alloys as a function of carrier concentration at three consecutive temperatures.

The room temperature Seebeck coefficient (S), electrical conductivity (σ), power factor (PF) (S2σ), and total thermal conductivity at their intrinsic carrier concentration are listed in Table IV, along with other available theoretical data. From the Table, it is clear that S is smaller than that of reported in the Ref. by PBE functional (which underestimates bandgap largely and hence overestimates carrier density) without SOC. But the electrical conductivity largely differs because it does not only depends on accurate bandgap but also on the accurate holes relaxation time calculation.

Table IV: Calculated room temperature thermoelectric parameters and available theoretical values (enclosed in the bracket).

| Parameters | TaSbRu | 25% Bi | 50% Bi |
|---|---|---|---|
| n (e+19 cm-3) | 1.39(10 [26])(1 [27]) | 4.3 | 4.9 |
| S (µV/K) | 176 (~235 [26]) (181 [27]) | 202 | 193 |
| σ (10^4 S/m) | 12.4 (16.3 [26]) (~65 [27]) | 9.9 | 7.8 |
| τ (fs) | 65.77 (37.52 [26]) (100 [27]) | 60.97 | 47.25 |
| κtot (W/m K) | 0.5 (35 [26]) | 0.42 | 0.34 |
| PF (mW/m K2) | 3.83 (~9 [26]) (21.29*) (2.8**) | 4.04 | 2.93 |

*PF (from wrong multiplication of S and σ) reported in fig. 5. (d) in Ref. [27]  ** Using cRTA ( τ=10 fs) [25]

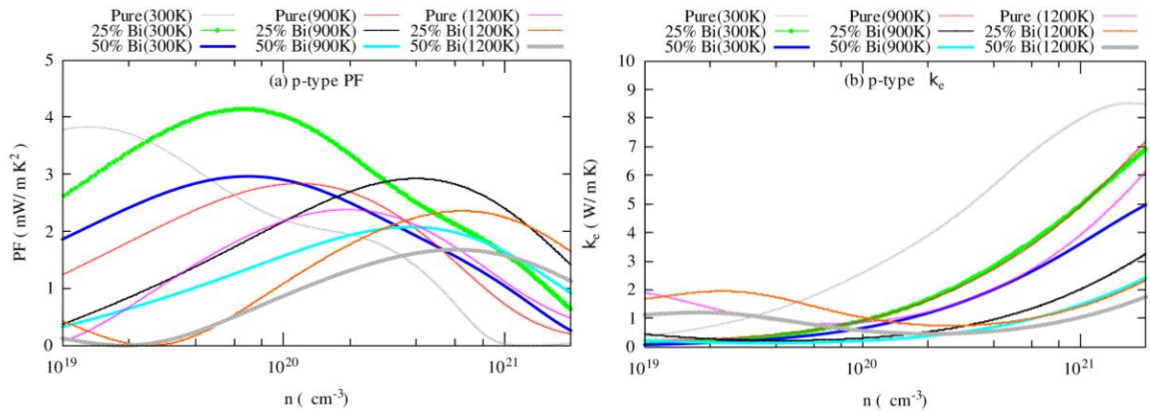

Fig. 8. (a) P-type power factor (PF) and (b) electronic part of the thermal conductivity of the studied compound and alloys as a function of carrier concentration at three consecutive temperatures.

Fig. 8. shows the computed power factor (PF) and the electronic part of the thermal conductivity (κe). Although the PF of pure TaSbRu is very high (3.83 mW/m K2) (much higher than that of

HfCoSb 0.05 mW/m K2), the electronic and lattice thermal conductivities are also high. Thus, the resultant thermoelectric performance would be negligibly small at this temperature. Upon Bi-substitution, the PF is substantially improved further at low temperatures but unfortunately reduced at high temperatures. Because the increase of both Bi concentration and temperatures intensifies the carrier scattering, such type of change is expected. However, the electronic part of the thermal conductivity is also reduced dramatically by Bi and temperatures. Therefore, the figure of merit would be optimum at high temperatures and Bi concentration. Note that Bi does not only causes the increase of PF at low temperature and vice versa but also it shifted the PF peak to the high carrier density, which is consistent with our projected DOS calculations.

### 3.4. Thermoelectric performance

By using the power factor (PF) and total thermal conductivity ($\kappa_e + \kappa_l$) as obtained in the previous section, our calculated thermoelectric figure of merit (ZT) at three consecutive temperatures as a function of carrier concentration is shown in Fig. 9. At room temperature, ZT is very small and even close to zero for pure TaSbRu. By the Bi substitution, the ZT is improved dramatically, as expected. It rises from 0.03 to 0.15 by 50% Bi at 300 K. At high temperatures, the effect of alloying is even higher. The slight increase of bandgap by Bi would be favorable to hinder the minority charge carriers from exciting intrinsically. This would minimize the hole-electron scattering.

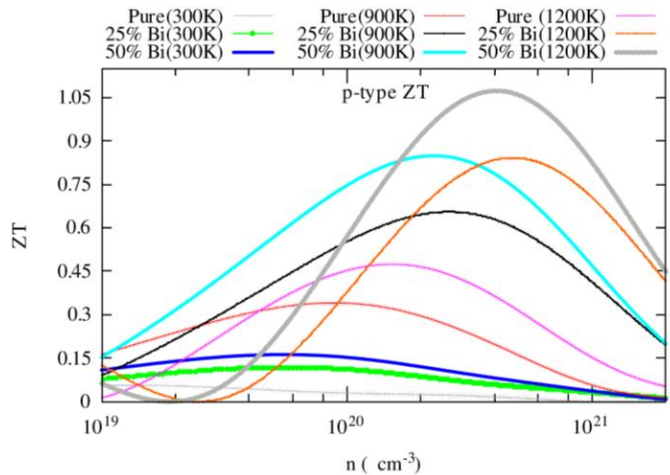

Fig. 9. P-type thermoelectric figure of merit (ZT) of the studied compound and alloys as a function of carrier concentration at three consecutive temperatures.

The computed ZT of pure TaSbRu is 0.45 at 1200 K, which is too small for practical applications. However, the ZT rises to ~1.1 from 0.45 at 1200 K by 50% Bi substitution, which is 2.45 times higher than that of a pure one. Such high ZT is suitable for thermoelectric device applications operating at high temperatures. A similar type of ZT improvement might be possible in other Sb-based half-Heusler compounds by Bi substitution. We hope that the experimentalist might be able to synthesize these alloys and verify our interesting and useful predictions of ZT improvement.

## 4. Conclusions

In summary, we have studied the effect of Bi-substitution on the structural stability and high-temperature thermoelectric performance of p-type half-Heusler TaSbRu. Interestingly, the substitution of Bi improves the dielectric properties of TaSbRu slightly and causes LO-TO splitting. The lattice dynamics calculations indicate that the studied compound and alloys are thermodynamically stable. The electronic structure calculations reveal that Sb has little contribution to the bandgap formation and all the systems have highly dispersive and degenerate valence bands (eightfold degenerate in TaSbRu and more than that in the alloys). The spin-orbit coupling splits both valence bands and conduction bands largely. The computed power factor of TaSbRu, an indirect bandgap semiconductor ($E_g$=0.8 eV by TB-mBJ+SOC), 3.8 mWm$^{-1}$K$^{-2}$ at 300K( comparable to Hf-doped FeNbSb (5.5) but 2.5 times smaller than that of obtained for pure TaSbRu considering energy independent relaxation time). This indicates the importance of energy-dependent hole relaxation time on the accurate description of power factor and hence thermoelectric properties. The substitutions of Bi on the Sb site do not cause significant change to the electronic structure but only a slight increase in the bandgap from 0.8 to 0.83 eV (within the computation accuracy) has been found. Although the Seebeck coefficient increases by Bi due to such bandgap change, the electrical conductivity, and hence, the power factor reduces to ~3 mWm$^{-1}$K$^{-2}$ at 300K (50% Bi). The substitution reduces the contribution of Ta to the lower energy phonons and on the other side, it increases Bi contribution. The anharmonicity increases significantly and hence, intensifies phonons scattering. Thus, the lattice thermal conductivity is reduced effectively to 5 from 20 W/m K at 300K. With the combination of high power factor and low thermal conductivity, the ZT value is improved dramatically, reaching 1.1 (50% Bi) at 1200 K from 0.45 (pure TaSbRu). Therefore, Our study suggests that the substitution of the heavier elements in Sb-based Half-Heusler compounds might be effectively reduced lattice thermal conductivity. Furthermore, TaSbRu is a promising material for high-temperature applications and we hope that experimentalists will synthesize the studied alloys soon.